# Shannon Entropy for Time-Varying Persistence of Cell Migration

Yanping Liu[1,*], Yang Jiao[2,*], Qihui Fan[3], Guoqiang Li[1], Jingru Yao[1], Gao Wang[1], Silong Lou[4], Guo Chen[1], Jianwei Shuai[5,†], Liyu Liu[1,†]

[1] *Chongqing Key Laboratory of Soft Condensed Matter Physics and Smart Materials, College of Physics, Chongqing University, Chongqing, 401331, China*
[2] *Materials Science and Engineering, Arizona State University, Tempe, Arizona 85287, USA*
[3] *Beijing National Laboratory for Condensed Matte Physics and CAS Key Laboratory of Soft Matter Physics, Institute of Physics, Chinese Academy of Sciences, Beijing 100190, China*
[4] *Department of Neurosurgery, Chongqing Cancer Hospital, Chongqing, 400030, China*
[5] *Department of Physics, Xiamen University, Xiamen, 361102, China*

[*] These authors contributed equally to this work.
[†] Corresponding authors: jianweishuai@xmu.edu.cn, lyliu@cqu.edu.cn

**ABSTRACT**

Cell migration, which can be significantly affected by intracellular signaling pathways (ICSP) and extracellular matrix (ECM), plays a crucial role in many physiological and pathological processes. The efficiency of cell migration, which is typically modeled as a persistent random walk (PRW), depends on two critical motility parameters, i.e., migration speed and persistence. It is generally very challenging to efficiently and accurately extract these key dynamics parameters from noisy experimental data. Here, we employ the normalized Shannon entropy to quantify the deviation of cell migration dynamics from that of diffusive/ballistic motion as well as to derive the persistence of cell migration based on the Fourier power spectrum of migration velocities. Moreover, we introduce the time-varying Shannon entropy based on the wavelet power spectrum of cellular dynamics and demonstrate its superior utility to characterize the time-dependent persistence of cell migration, which is typically resulted from complex and time-varying intra or extracellular mechanisms. We employ our approach to analyze trajectory data of *in vitro* cell migration regulated by distinct intracellular and extracellular mechanisms, exhibiting a rich spectrum of dynamic characteristics. Our analysis indicates that the combination of Shannon entropy and wavelet transform offers a simple and efficient tool to estimate the persistence of cell migration, which may also reflect the real-time effects of ICSP-ECM to some extent.

1. Introduction

Cell migration [1] is a ubiquitous and crucial phenomenon that is found in many physiological processes including neural system development [2], wound healing [3,4], and immunological responses [5]. Eukaryotic cell migration [6] is a complex behavior involving various cellular and sub-cellular level events, and is strictly regulated by intracellular signaling pathways (ICSP) [7,8] and extracellular matrix (ECM) [9-12]. Many human diseases are associated with ill-regulated cell migration, among which cancer metastasis is the most representative and fatal case [13,14].

In order to get insights into the cell migration in ECM, a number of *in vitro* experiments attempting to mimic various aspects of realistic *in vivo* environments have been designed and carried out in recent years. For instance, stiffer substrates usually increase the persistence of cell migration, while the soft substrates typically lead to apparently more random motions [15]. This unusual behavior depending on the substrate stiffness is typically referred to as *durotaxis*, a mechanism that can regulate many pathological processes when combining with mechanical strains [16]. Besides the substrate properties, it has been shown that nanoscale topographic features in ECM, coupled with the effective stiffness of cells, can guide persistent migration, which is called as *topotaxis* [17]. In addition, it was found that the heterogeneous topology of ECM can boost cell invasion into a 3D funnel-like matrigel interface in a micro-fabricated biochip [18]. Aligned fibers can also facilitate the migration of MDA-MB-231 breast cancer cells into rigid matrigel in a constructed collagen I–matrigel microenvironment [19].

In order to phenomenologically investigate and quantify the rich spectrum of cellular migratory behaviors, many motility models have been developed [20]. For instance, amoeba performs a special random walk, which can increase the probability of find a target in complex microenvironment [21]. Similarly, CD8 (+) T cell adopts a strategy known as generalized Lévy walk, which contributes to finding rare targets [22]. Among these models, the persistent random walk (PRW) [23-25] is one of the most representative and commonly used model, which incorporates the memory of cell to the past velocities based on Brownian motion [26], and can be derived from the following Langevin equation [27]

$$\frac{d\vec{v}}{dt} = -\frac{\vec{v}}{P} + \frac{S}{\sqrt{P}} \cdot \tilde{w}, \tag{1}$$

where $\vec{v}$ is the migration velocity, P the persistence time, S the averaged migration speed and $\tilde{w}$ the random vector of a Wiener process [28]. Here, both of P and S are called "motility parameter" and together determine the cell migration capability.

In studying cell behaviors in complex microenvironment, how to efficiently and accurately quantify cell migration capability has become a crucial issue. Thus, a number of works have been implemented to derive some estimators that can characterize cell migration capability [29,30]. For instance, the diffusion coefficient of one cell can be obtained from a time-lapse recorded trajectory by developing optimal methods [31,32]. In addition to the diffusion coefficient, direction autocorrelation and other parameters are computed to analyze cell migration in two dimensions by running a computer program, DiPer [33]. Moreover, motility parameters (e.g., persistence time and migration speed) can be derived from the fittings to mean squared displacement (MSD) [33-35], velocity autocovariance function (VAC) [24] and Fourier power spectrum (FPS) [34].

Different from the aforementioned cases in which the dynamical parameters and cellular/environmental properties do not change with time, cell migration can be significantly affected by many factors [1], and thus exhibits time-varying migratory capability and behaviors. For example, the tissue with cultured polystyrene surface can increase a motion coefficient over time, while the motion coefficient remained relatively constant for tissue with untreated polystyrene plates [36]. Accordingly, the time-dependent parameters (persistence and activity) were also derived from time-lapse recorded trajectories with a Bayesian method [37]. Note that all parameters derived from the aforementioned methods can characterize cell migration capability, but may show inaccuracy to some extent, because of (i) the fact that some novel cell behaviors are not described well by present theoretical models [20], and (ii) these methods mainly include fittings (goodness of fit, $R^2$) based on the theoretical models and are inevitably associated with numerical errors.

In this work, we introduce the time-varying Shannon entropy based on the wavelet power spectrum obtained by performing wavelet transform of migration velocities and demonstrate its superior utility to characterize the persistence of cell migration. In order to establish the framework and verify its accuracy, we first employ the time-varying persistent random walk model to simulate cell migration with time-varying characteristics to generate synthetic testing data and compute the Fourier power spectrum of migration velocities. Second, we analyze the effects of individual parameters on Fourier power spectra and introduce normalized Shannon entropy to characterize the persistence of cell migration, inspired by changes in power spectra. Then, we perform the wavelet transform of migration velocities to obtain wavelet power spectrum, which is employed to derive the time-varying Shannon entropy that can reflect the time-varying persistence. Finally, two indicators have been further defined to estimate the overall persistence of a cell population.

With the method verified, we employ our methods to analyze trajectory data of *in vitro* cell migration regulated by distinct intracellular and extracellular mechanisms, exhibiting a rich spectrum of dynamic characteristics and persistence. In particular, the Shannon entropy is normalized to the interval [0, 1], which can be viewed as an estimator to quantify how much the cell migration dynamics deviates from that of a Brownian motion or ballistic motion. Our analysis indicates that the combination of Shannon entropy and wavelet transform offers an efficient tool to estimate the persistence of cell migration, which may also reflect the real-time effects of ICSP-ECM to some extent.

The rest of the paper is organized as follows: In Sec. 2, we simulate cell migration based on the time-varying persistent random walk model and compute Fourier power spectrum of cell migration velocities. By analyzing the changes in power spectral values, we further introduce Shannon entropy to reflect the persistence of cell migration. In Sec. 3, we perform the wavelet transform of cell migration velocities and obtain wavelet power spectrum, which is combined with Shannon entropy to quantify the time-varying persistence of cell migration. In Sec. 4, we compute Shannon entropies of *in vitro* cell migrations regulated by distinct intracellular and extracellular mechanisms, which verifies the efficiency and practicality of approach developed. In Sec. 5, we provide concluding remarks.

## 2. Shannon entropy characterizing the persistence of cell migration

In this section, we provide the framework for Shannon entropy based on the analysis of persistent cell migration. We first simulate cell migration with the time-varying persistent random walk (TPRW) model. Second, the corresponding Fourier power spectrum of migration velocities is calculated based on the Wiener-Khinchin theorem. Inspired by the changes in power spectral values, we finally introduce the Shannon entropy to characterize the changes and further quantify the persistence of cell migration.

*2.1. Cell migration with time-varying characteristics*

Usually, cell migration exhibits the time-dependent characteristics because of the effects of many factors [1], among which the ECM is the most representative one. Here, the "time-dependent" means that cell migration capability will vary with time, and cannot be described well by Langevin equation with constant parameters P and S [27]. Thus, It is necessary to generalize the Langevin equation with time-varying parameters, i.e., P(t) and S(t), and both of them together quantify cell migration capability. Further, the persistent random walk (PRW) model derived from the Langevin equation with constant parameters will be generalized to the time-varying persistent random walk (TPRW) model, see detail simulations in subsection 2.2.

Considering the time-varying characteristics of cell migration, we first construct function P(t) based on a linear variation [37], which is of following form

$$P(t) = K_P \cdot t + P_0, \quad (2)$$

where $K_P$ is the changing rate of persistence time, and $P_0$ is initial value at t = 0 min. The case of $K_P$ = 0 denotes that cell migration capability is a constant function of time, and corresponds to the PRW model. It was recently reported that there is a correlation between the persistence time P and migration speed S, which can be well fitted by a simple exponential curve before saturating at larger speed [38], as follows

$$P=Ae^{\lambda S}, \tag{3}$$

where A and λ are constant, which are mainly determined by cell type and ECM. Thus, migration speed S(t) can be derived when the constants in Eq. (3) are determined. Note that A is always assigned as 1 in this work.

*2.2. Numerical simulations of cell migration based on TPRW model*

Using a similar procedure for the PRW model [30,39], cell migration described by TPRW model can be performed in computer simulations. We will mainly focus on 2D migrations in the ensuing discussions and the generalization to 3D is straight forward. In particular, the 2D position for one cell at each time step can be readily computed using the following functions

$$x(t+\Delta t)=x(t)+\Delta x(t,\Delta t), \tag{4}$$

$$y(t+\Delta t)=y(t)+\Delta y(t,\Delta t), \tag{5}$$

where Δx and Δy are displacements for a given time interval Δt, and they are written as

$$\Delta x(t,\Delta t)=\alpha(t)\cdot \Delta x(t-\Delta t,\Delta t)+F(t)\cdot \tilde{W}, \tag{6}$$

$$\Delta y(t,\Delta t)=\alpha(t)\cdot \Delta y(t-\Delta t,\Delta t)+F(t)\cdot \tilde{W}, \tag{7}$$

where $\alpha(t) = 1-\Delta t/P(t)$ showing the memory of one cell to the past velocities, $F(t) = [S(t)^2\cdot \Delta t^3/P(t)]^{1/2}$ quantifying the amplitudes of Gaussian white noise $\tilde{W}$, which is also called intrinsic noise in cell dynamics. On the one hand, it is evident that the parameter α approaches to 1 and the F does to 0 when P tends to infinity for a given time step, which indicates that cell will migrate along a fixed direction without turning. Similarly, the α approaches to 0 and the F does to S(t)·Δt when P tends to Δt, which means that the direction of cell migration cannot be predicted, corresponding to the normal Brownian motion. On the other hand, the α always lies in the interval [0, 1], thus the displacements will gradually decrease to 0 with time lapsing without taking into account the contribution of F. Thus, we can view the α and F as slow-down and speed-up factors, respectively.

After generating cell migration trajectories, we further add positioning errors to the simulated data to mimic the effects of experimental observations, as follows

$$\hat{x}(t)=x(t)+\sigma_{pos}\cdot \tilde{W}, \tag{8}$$

$$\hat{y}(t)=y(t)+\sigma_{pos}\cdot \tilde{W}, \tag{9}$$

where $\sigma_{pos}$ is positioning error and assigned as 0.01 μm [24] in this work.

*2.3. Fourier power spectrum of cell migration velocities*

In order to get insights into cell migration, we define a set of parameters $P_0$ = 8.0 min, $K_P$ = 0, $S_0$ = 0.5 μm/min (A = 1 and λ = 4.16), based on the Eqs. (2-3). Here, motility parameters are non-varying with time, which reflect the possible stable effects of ICSP-ECM on cell motility, to some extent. Therefore, cell migration underlying these parameters can be described by PRW model. Next, we follow the procedure above [cf. Eqs. (4-9)] to simulate 200 cell migration trajectories and each trajectory contains 4800 + 1 (N + 1) frames, i.e., total time T = 960 min, one of which is plotted in Fig. 1(a). Further, the velocity components in x and y axes are also obtained from cell positions, based on the displacements for time step Δt = 0.2 min, as shown in Fig. 1(b). It seems like that the velocity components keep stable with time lapsing.

In general, one can first calculate any physical quantity for figuring out the properties concerning cell migration, e.g., mean squared displacement (MSD), velocity autocovariance function (VAC) or Fourier power spectrum (FPS). Second, the motility parameters can be derived from the fittings to the calculated quantities with the corresponding theoretical formula [33,34,39]. Finally, the resulting parameters can be analyzed by statistical methods to extract the desirable properties and characteristics. In this work, we follow these procedures but without numerical fittings, to quantify the persistence of cell migration. We introduce the VAC for individual cell migration velocities, which is given as

$$\text{VAC}(t_j) = \langle \vec{v}_i \cdot \vec{v}_{i+j} \rangle = \frac{1}{N-j-1} \sum_{g=1}^{N-j} \left( \vec{v}_g - \frac{1}{N-j} \sum_h^{N-j} \vec{v}_h \right) \cdot \left( \vec{v}_{g+j} - \frac{1}{N-j} \sum_{h=j+1}^{N} \vec{v}_h \right), \tag{10}$$

where $t_j = j \cdot \Delta t$, N is the total number of migration velocities per trajectory, and lower case letters (i, j, g and h) are time indexes. The resulting VAC for individual trajectories is shown in Fig. 1(c). It is obvious that the VAC follows a nonlinear exponential decay in linear-log axes, which is the consequence of finite migration data. It is well known that the VAC for an Ornstein-Uhlenbeck (OU) process [40] obeys a linear exponential decay, thus the nonlinearity in Fig. 1(c) will transit to linearity when considering a large number of cell migration data.

Although the VAC is a classical and widely used approach to analyze the properties of cell migration, especially the persistence, it still has limitations (see Sec. 1). For instance, it could not return reliable errors on the fitted motility parameters, because of the correlations between velocities in time domain [34]. In order to eliminate the correlations, a novel quantity, Fourier power spectrum (FPS), has been introduced based on the Wiener-Khinchin (WK) theorem [41,42], which states that "the power spectrum of any generalized stationary random process is the Fourier transform of its autocovariance function." According to this theorem, the Fourier transform of migration velocities is given as follows

$$\hat{\vec{v}}_k = \Delta t \sum_{j=1}^{N} e^{i2\pi f_k t_j} \cdot \vec{v}_j = \Delta t \sum_{j=1}^{N} e^{i2\pi kj/N} \cdot \vec{v}_j, \tag{11}$$

and similarly the Fourier transform of velocity autocovariance function is given as

$$\widehat{\text{VAC}}(f_k) = \Delta t \sum_{j=1}^{N} e^{i2\pi f_k t_j} \cdot \text{VAC}(t_j) = \Delta t \sum_{j=1}^{N} e^{i2\pi kj/N} \cdot \text{VAC}(t_j), \tag{12}$$

finally, the Fourier power spectrum is

$$\text{FPS}(f_k) = \left\langle \left| \hat{\vec{v}}_k \right|^2 \right\rangle \Big/ T = \frac{(\Delta t)^2}{T} \sum_{j_1=1}^{N} \sum_{j_2=1}^{N} e^{i2\pi f_k (t_{j_1} - t_{j_2})} \cdot \langle \vec{v}_{j_1} \cdot \vec{v}_{j_2} \rangle = \Delta t \sum_j e^{i2\pi f_k t_j} \cdot \text{VAC}(t_j) = \widehat{\text{VAC}}(f_k), \tag{13}$$

where $f_k = k/T$ is Fourier frequency, k is frequency index, $T = N \cdot \Delta t$ is total time for recording individual trajectories, $j_1$, $j_2$ are time indexes and the symbol "^" denotes Fourier transformed quantity. Thus, the corresponding FPS can be easily computed and then plotted in log-log axes, as shown in Fig. 1(d).

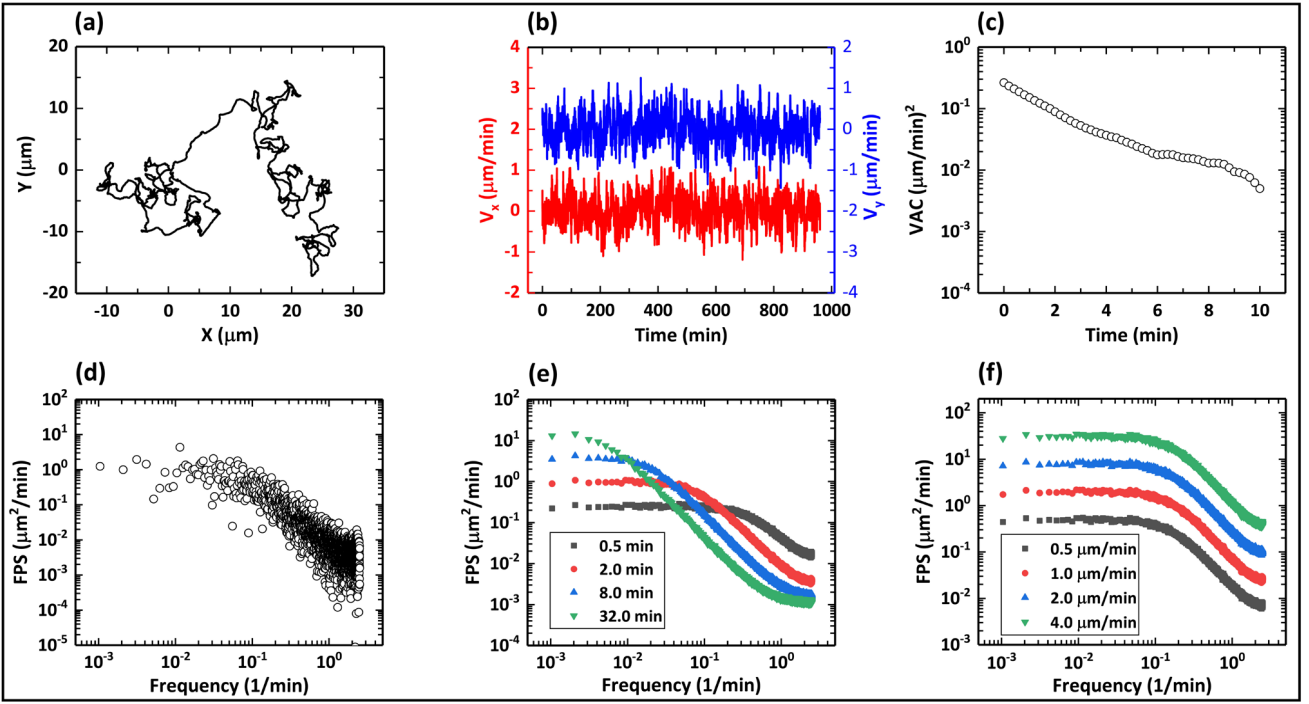

**Fig. 1.** Fourier power spectrum of cell migration velocities. (a) Individual cell migration trajectories simulated by TPRW model. (b) Velocity components in x and y axes. The red line corresponds to components in x axis, while the blue does to those in y axis. (c) Velocity autocovariance function (VAC) in linear-log axes for individual cells. (d) Fourier power spectrum (FPS) in log-log axes for individual cells corresponding to the VAC in (c). (e) The effects of persistence time P on Fourier power spectrum. (f) The effects of migration speed S on Fourier power spectrum.

Next, we continue to analyze the effects of persistence time P and migration speed S on FPS via *control variable*, thus several sets of motility parameters have been defined based on the Eqs. (2-3), as given in Table 1. There, all motility parameters are initial values at t = 0 min, and $K_P$ equals to 0. The left part in the table aims to study the effects of persistence time on FPS, while the right does to study those of migration speed.

**Table 1** Prescribed motility parameters in TPRW model

| Increasing persistence time P (min) | | | | Increasing migration speed S (µm/min) | | | |
|---|---|---|---|---|---|---|---|
| Code | $P_0$ | $\lambda$ | $S_0$ | Code | $P_0$ | $\lambda$ | $S_0$ |
| P1 | 0.5 | -1.39 | 0.5 | S1 | 1.0 | 0 | 0.5 |
| P2 | 2.0 | 1.39 | 0.5 | S2 | 1.0 | 0 | 1.0 |
| P3 | 8.0 | 4.16 | 0.5 | S3 | 1.0 | 0 | 2.0 |
| P4 | 32.0 | 6.93 | 0.5 | S4 | 1.0 | 0 | 4.0 |

The corresponding results are shown in Figs. 1(e-f). It is clear that the FPS behave differently in the four cases of persistence time P [see Fig. 1(e)], i.e., the horizontal regions (e.g., 0.001 ~ 0.2 /min for the code P1) gradually get narrower while the decay regions become wider (e.g., 0.2 ~ 2.5 /min for the code P1), and the horizontal regions correspond to greater power spectral values, as the persistence time P increases. Thus, the FPS will transit from the horizontal curve to a sharp decaying curve, which correspond to Brownian motion and ballistic motion, respectively. In other words, the power spectral values in a given frequency domain will change from the uniform to non-uniform. In contrast, Fig. 1(f) indicates that the increasing of S only increases the amplitudes of FPS, instead of changing the decay rates. When the motility parameters are different from the counterparts for other cases, the corresponding FPS will exhibit more complex changes, which can be directly quantified by the Lorentzian power spectrum [30,34].

*2.4. Shannon entropy*

Inspired by the changes in Fourier power spectra (FPS) [see Figs. 1(e-f)], we introduce Shannon entropy to analyze the information contained in FPS, especially the persistence. Entropy is an extensively used concept in Thermodynamics, which is typically to describe the degree of disorder or randomness in the states of molecules. It was not until 1948 that it was introduced to describe the uncertainty in information source, by C. E. Shannon [43]. Therefore, it is also referred to as the *Shannon entropy* when related to information theory.

Suppose these is a set of possible events with probabilities of occurrence $p_1, p_2, ..., p_n$, the Shannon entropy is given as

$$H=-\sum_{i=1}^{n} p_i \cdot \log2(p_i), \qquad (14)$$

where n is the total number of the events and $p_i$ represents the probability of each event. Note that the Shannon entropy is more commonly denoted by capital letter "H" in literatures. The H not only measures how much "choice" is involved in the selection of the event, but also how uncertain the outcome can be. In order to illustrate the relationship between the probabilities of events and H, we consider a case of two possibilities with probabilities p and q, whose H is written as

$$H=-\left(p \cdot \log2(p)+q \cdot \log2(q)\right), \qquad (15)$$

where p lies in the interval [0, 1], and q = 1 – p. The result is plotted in Fig. 2(a), which shows that the H first increases and then decreases as the probability p increases. Besides, it is symmetry about p = 0.5, which means that the H reaches maximum 1 bit when the probabilities of events are same, and we are not completely certain of outcome. Inversely, when p = 0 or 1, H will reach a minimum 0 bit, thus we are completely certain of outcome. Otherwise, H lies in the interval (0, 1).

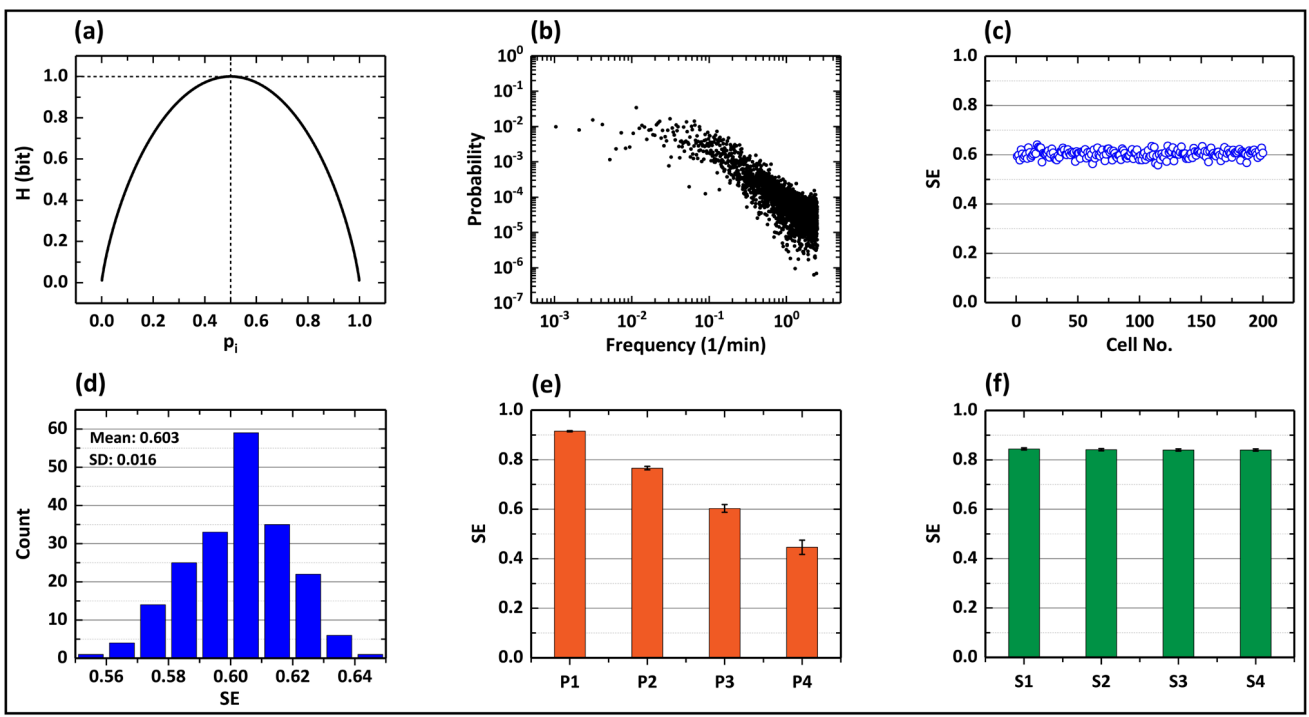

**Fig. 2.** Shannon entropy characterizing the persistence of cell migration. (a) Shannon entropy (H) as a function of probability in the case of two possibilities. (b) The probability as a function of frequency obtained from the normalization to Fourier power spectrum in Fig. 1(d). (c) Shannon entropies (SEs) for a given cell population. (d) Distribution of Shannon entropies (SEs) for a given cell population with mean and standard deviation (SD). (e) Shannon entropies (SEs) in the cases of four different persistence times. (d) Shannon entropies (SEs) in the cases of four different migration speeds. Note that the error bas are standard deviation (SD).

Inspired by the analysis of H for the case of two possibilities, we first normalize the individual power spectral values in Fig. 1(d) by dividing the sum of individual power spectral values in the entire frequency domain, as shown in Fig. 2(b). The normalized spectral values and frequencies can be viewed as probabilities and events in Eq. (14), respectively, which represents the probability of occurrence for each frequency. Although the shape of normalized result is identical with that of FPS, it possesses completely different physical interpretations.

Following the procedure above, we calculate the H for every cell using Eq. (14). Further, we normalize all Shannon entropies (Hs) by dividing the maximal H corresponding to the case with same occurrence probabilities for all events, which excludes the effects of the total number of frequency on Hs. After normalization of all Hs, all normalized Hs are dimensionless and locate in the interval [0, 1] [see Fig. 2(c)], which makes it convenience to compare with other cases. Here, we employ the abbreviation *SE* to denote the normalized H for avoiding confusion in this work. Figure 2(c) clearly shows that all SE values fluctuate around 0.6 with small derivations, which reflects the similarities between cells in a given cell population. Moreover, the distribution of all SEs in Fig. 2(d) vividly exhibits the mean (~ 0.603) and SD (~ 0.016) of a cell population in a more straightforward manner. Because the cell population here is modeled based on the same motility parameters, these is no big differences between all SEs. Similarly, we computed the overall-averaged SE for each case with 200 cells listed in Table 1. The details are shown in Figs. 2(e-f). In Fig. 2(e), the overall-averaged SE gradually decreases when the persistence time P increases, which indicates that the more persistent the cell migration, the smaller the corresponding SE. However, the overall-averaged SE in Fig. 2(f) is almost identical with each other, which illustrates that migration speed S will not affect the persistence of cell migration.

## 3. The time-varying Shannon entropy for cell dynamics

In this section, we introduce wavelet transform to compute wavelet power spectrum of cell migration velocities, and further derive the time-dependent Shannon entropy (SE), which can accurately quantify the time-varying persistence of cell migration.

### 3.1. The time-varying motility parameters with individual differences

Different from the cases with constant motility parameters in Sec. 2, the function P(t) is constructed again, based on two Gaussian distributions, i.e., $N_1(2.0, 0.1^2)$ and $N_2(20.0, 1.0^2)$. The persistence times at initial P(0) and final P(T) moments are determined by the distributions $N_1$ and $N_2$, respectively. The persistence time still obeys a linear function [cf. Eq. (2)]. Furthermore, the migration speed can be computed using Eq. (3) when both of A and $\lambda$ are equal to 1. So far, the motility parameters for a cell population have been constructed, which exhibit three characteristics: (i) both of them vary with time and the former changes linearly while the latter exhibits nonlinear behaviors, (ii) both of them increase with time, and together reflect the *enhance* of migration capability, and (iii) the motility parameters for every cell are different from the corresponding parameters of another cells, reflecting the heterogeneities among the cell population.

Figure 3(a) shows how the motility parameters change over time for one cell, and the corresponding trajectory is plotted in Fig. 3(b). The velocity components in Fig. 3(c) exhibit greater fluctuations when comparing with those in Fig. 1(b), which may illustrate the instability of cell migration to some extent.

*3.2. Wavelet power spectrum of cell migration velocities*

Since cell migration is affected by many factors, it is necessary to study the time-varying characteristics encoded in cell migration trajectories. Although the combination of Fourier power spectrum (FPS) and Shannon entropy (SE) used above is able to capture the degree of persistence for each trajectory, it is incapable of obtaining time-dependent information [44]. In order to address this problem for a time series, researchers have developed two methods, namely windowed Fourier transform (WFT) and wavelet transform (WT).

The WFT is an analytical method to extract time-frequency information from a time series, which performs the Fourier transform on a sliding segment of a constant time interval from a time series. Here, the segment can be windowed with an arbitrary function, e.g., a boxcar or a Gaussian window. Although the WFT shows the ability of extracting time-frequency information, it still cannot avoid several deficiencies. For instance, (i) it takes efforts to determine the most appropriate window size; (ii) the aliasing of high and low frequency components may not fall within the frequency range of the window; (iii) the frequencies corresponding to the segment must be analyzed at each time step, regardless of the window size or the dominant frequencies. These deficiencies make the WFT inaccurate and inefficient under certain situations, as discussed by Kaiser *et al.* [45], Torrence *et al.* [46], and Daubechies [47].

Different from the WFT, the window size of wavelet transform (WT) can vary with frequency, which allows one to analyze the time-frequency characteristics of a time series [46-48]. The WT was originally employed by Morlet *et al.* to analyze seismic signals in the early 1980s [49,50], and later formalized by Goupillaud and Grossmann *et al.* [51,52]. Due to the better performance in studying the non-stationary and infinitely correlated processes, the WT has become an influential tool. For instance, the wavelet coefficients of fractional Brownian motion are stationary and uncorrelated [53]. Note that the WT here includes discrete WT and continuous WT, the latter is employed in this work. For a given time series $v_j$, the continuous WT [46] is performed by computing the convolution of $v_{j'}$ with a scaled and the translated version of the wavelet function $\psi_0^*$, which is written as

$$W_j(s) = \sum_{j'=0}^{N-1} v_{j'} \cdot \psi_0^* \left[ \frac{(j'-j) \cdot \Delta t}{s} \right], \tag{16}$$

where the asterisk "*" represents the complex conjugate, and s is wavelet scale relating to Fourier frequency. Moreover, Morlet wavelet is employed here as wavelet function, which consists of a plane wave modulated by a Gaussian

$$\psi_0(\eta) = \pi^{-1/4} \cdot e^{i\omega_0 \eta} \cdot e^{-\eta^2/2}, \tag{17}$$

where $\omega_0 = 6$ is a non-dimensional frequency and satisfies the admissibility condition [54]. Note that the Morlet used is a complex function, thus the final results $W_j(s)$ is also complex. Further, we have access easily to the corresponding real part and imaginary part, both of them together are used to compute the wavelet power spectrum (WPS) [46,55], i.e., $|W_j(s)|^2$.

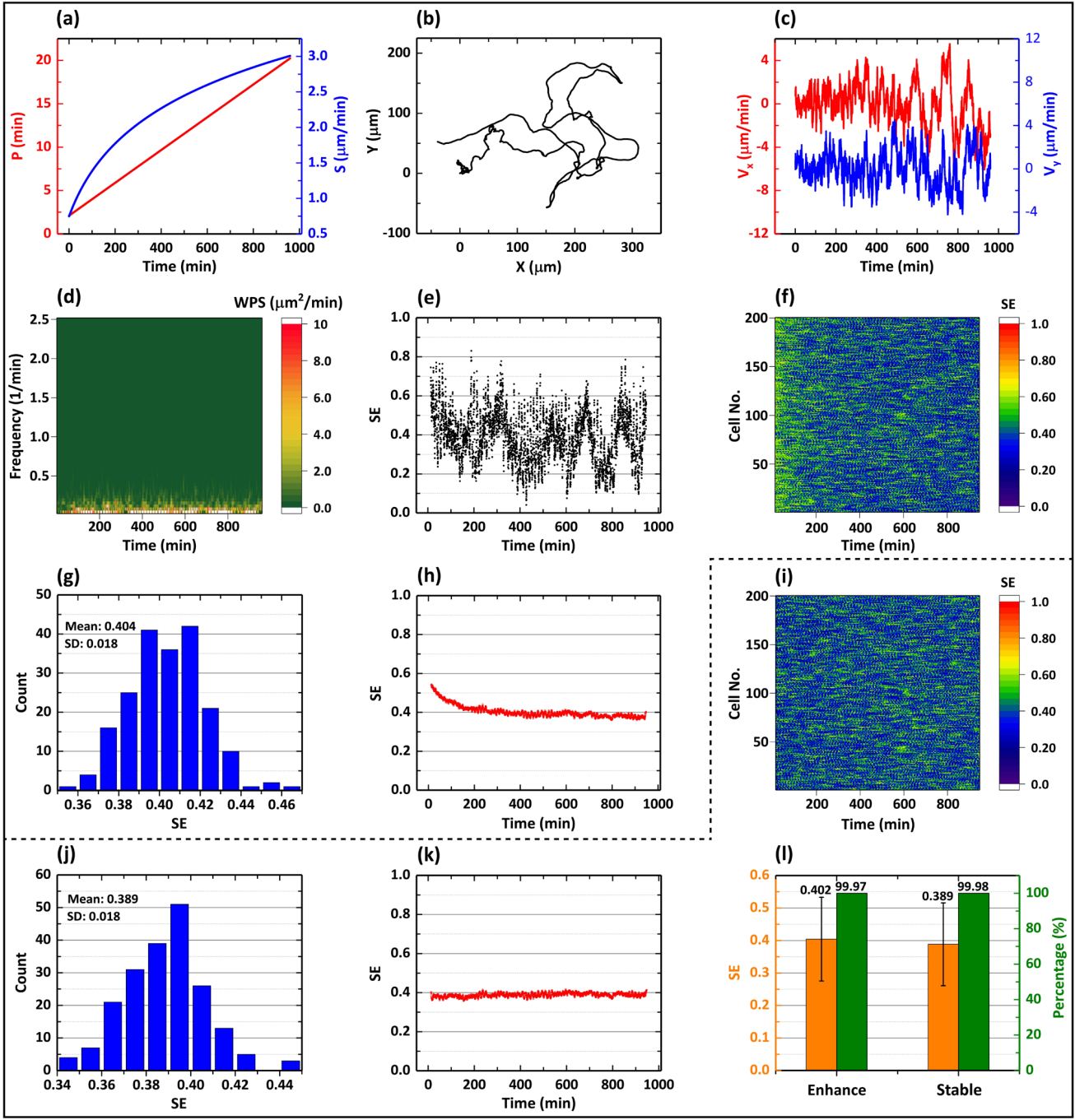

**Fig. 3.** The time-varying Shannon entropy of a cell population based on wavelet transform. (a) Motility parameters as functions of time. The red line indicates the persistence time P, while the blue line does the migration speed S. (b) Individual cell trajectories simulated by TPRW model based on the motility parameters in (a). (c) Velocity components in x and y axes. The red line denotes the components in x axis, while the blue does that in y axis. (d) Wavelet power spectrum (WPS) of individual cell migration velocities. The different colors denote the power spectral values in time-frequency domain. (e) The time-dependent Shannon entropies (SEs) for individual cells. (f) The time-dependent Shannon entropies (SEs) for a cell population. The different colors denote the values of Shannon entropy. (g) Distribution of the time-averaged Shannon entropies (SEs) for a cell population with 200 cells. (h) The cell-averaged Shannon entropy (SE) as a function of time. (i-k) The captions are same as those for (f-h), but the results are corresponding to the case with temporally non-varying motility parameters. (l) The overall-averaged Shannon entropy (SE) and the percentage of the entropies less than 0.8. The orange bars denote the average SE,

while the green do the percentage. The error bars are SD.

So far, we can analyze the characteristics of cell migration in time-frequency domain following the procedure above, as seen in Fig. 3(d). The horizontal axis corresponds to the total time for recording cell trajectory, while the vertical axis corresponds to the Fourier frequency. The legend in the right of the plot shows the power spectral values represented by different colors. Here, the power spectral values along the frequency axis is called local wavelet power spectrum (local WPS), and *local* means each moment of the total time. It is evident that the local WPS in the interval (0 ~ 0.5 /min) are greater than those in other ranges (0.5 ~ 2.5 /min), which roughly exhibits the persistence or correlations of cell migration velocities. When averaging the local WPS along the time axis, we can obtain time-averaged power spectrum, which is also called global wavelet power spectrum (global WPS). Torrence *et al*. [46] validated that the local WPS is identical with the FPS of an OU process, and the global WPS tends to approximate the FPS, thus it is reasonable to fit WPS with Lorentzian power spectrum [34].

Actually, edge effects will occur in the beginning and end of the WPS because of the finite-length time series, which is also called *cone of influence* (COI) [46]. Since the COI will greatly affect the true information, we exclude the region affected by COI in this work and the remaining region [see Fig. 3(a)] can exhibit the true characteristics of cell migration.

*3.3. The time-varying Shannon entropy*

Based on the wavelet power spectrum (WPS) in Fig. 3(d), we normalize the local WPS to obtain probabilities and compute the Shannon entropy (SE) for each moment, as shown in Fig. 3(e). Figure 3(e) shows the SE as a function of time, which also directly reflects the time-dependent persistence of cell migration [see Fig. 3(a)]. It is clear that the time-varying SE is not as smooth as that of P in Fig. 3(a), this difference is mainly caused by the intrinsic noises contained in cell dynamics [cf. Eq. (1)]. We continue to compute the time-varying SEs of a given cell population with 200 cells, and which are stacked along the vertical axis, as shown in Fig. 3(f). From the overall view, the time-varying SEs gradually decrease with time lapsing, which indicates that all persistence for this cell population gradually increase and this tendency is consistent with that of P in Fig. 3(a).

When averaging the time-varying SEs of this cell population along the time axis, we obtain the time-averaged SE for every cell, which quantifies the average persistence in the process of migration. The corresponding distribution is exhibited in Fig. 3(g), with mean (~ 0.404) and SD (~ 0.018) showing overall characteristics for this cell population, which can be easily used to compare with other cell populations. Similarly, we obtain the cell-averaged SE by averaging the time-varying SEs of this cell population along the vertical axis, as shown in Fig. 3(h). Different from the time-varying SE of individual cells [see Fig. 3(e)], the cell-averaged SE looks smoother, because the fluctuations caused by intrinsic noises have been averaged. Moreover, the gradually decrease of the cell-averaged SE illustrates that the persistence of cell migration gradually increases, which conforms the tendency of P in Fig. 3(a). Thus, the cell-averaged SE is better to reflect the overall changes against time in persistence of cell population, and also reflects the effects of ICSP-ECM on cell population, to some extent.

In order to validate the efficiency of our approach, we simulate 200 cell migration trajectories again with the same procedure as the case of enhanced migration capability [see Figs. 3(a-h)]. The only difference is that both of the initial and final persistence times are obtained from the Gaussian distribution $N_3(11.0, 0.1^2)$, and this case is referred as *stable*. The stable case means that cell migration capability almost does not change with time. The results are shown in Figs. 3(i-k). It is clear that the time-varying SEs of a cell population in Fig. 3(i) look more homogeneous, the time-averaged SE in Fig. 3(j) has a smaller mean 0.389 (SD: 0.018) and the cell-averaged SE in Fig. 3(k) exhibits a more stable tendency when compared with the results in Figs. 3(f-h).

In order to better compare the persistence of cell populations corresponding to different prescribed motility parameters (or cell populations in different microenvironment systems), we introduce two indicators based on the time-varying SEs of cell population, i.e., average of all SEs and percentage of the SEs less than 0.8. Figure 3(l) shows the comparisons of the two cell populations. The results indicate the average SE (0.389) for *stable* system is less than that (0.404) for *enhance* system, and the percentage (99.98%) of the former is greater than that (99.97%) of the latter, which means that cell migration in *stable* system is more persistent than that in *enhance* system. Note that the results here are determined by the prescribed motility parameters, which is mainly used to illustrate the usage of the two indicators introduced.

**4. Modeling and quantifying *in vitro* cell migration via Shannon entropy**

In previous sections, we have developed and verified a framework to derive the time-varying Shannon entropy for a cell population, which reflects the time-varying persistence of cell migration based on synthetic data generated from simulations. In order to further illustrate the utility of our framework, we employ methods to analyze trajectory data of *in vitro* cell migration regulated by distinct intracellular and extracellular mechanisms, exhibiting a rich spectrum of dynamic characteristics and persistence.

*4.1. Migration regulated by intracellular signaling pathway: Arpin proteins*

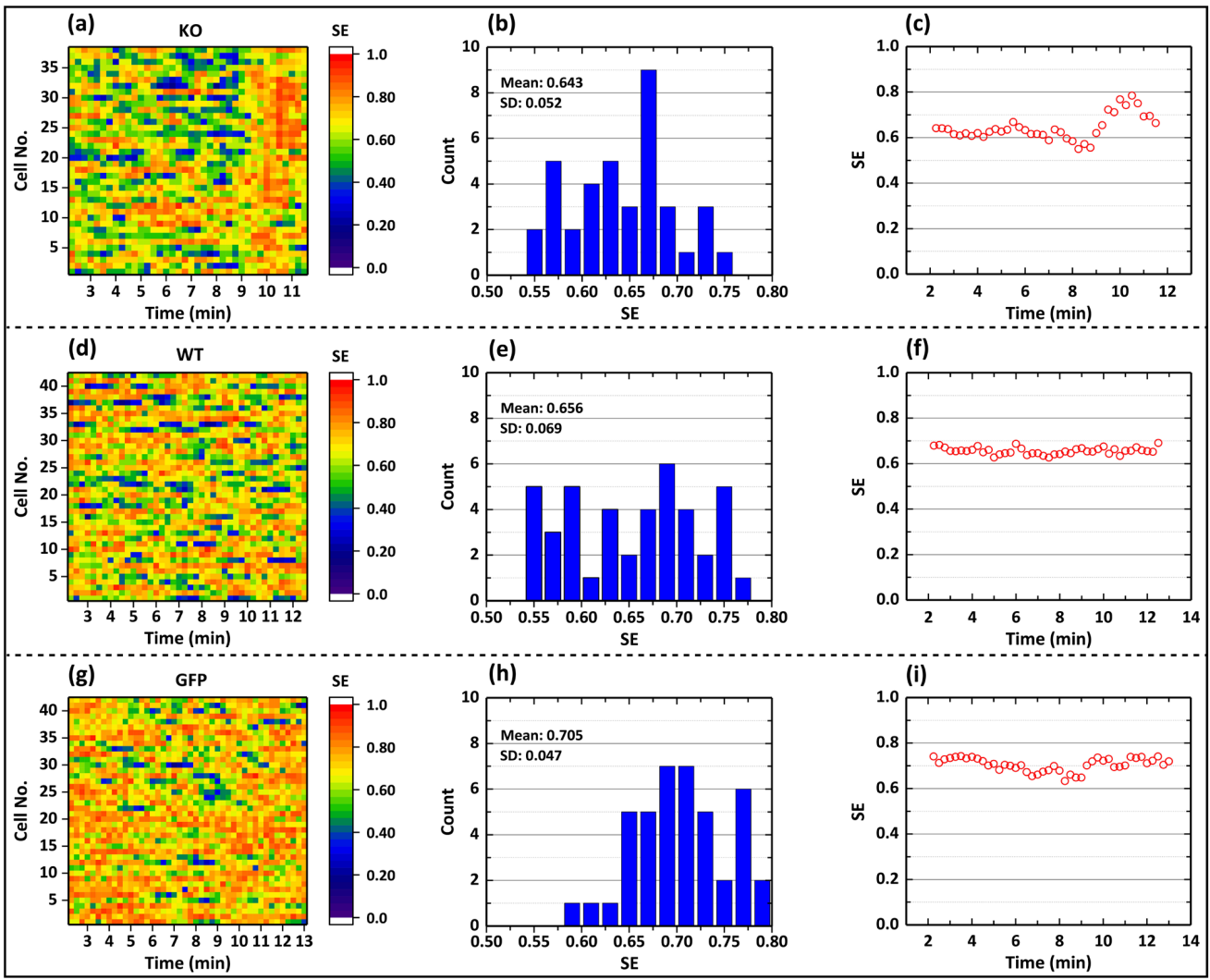

**Fig. 4.** Shannon entropy of the automatically tracked *D. discoideum* cells. The captions are same as those in Fig. 3(g-h), but the results corresponds to Arpin knockout (KO) (a-c), wild-type (WT) (d-f) and GFP-Arpin rescue amoeba (GFP)

(g-i), respectively. Here, the time interval between frames is 15 s and the number of cells are 38, 42 and 42 for KO, WT and GFP. All the experiment data is obtained from Ref. [33].

We first analyze migratory dynamics regulated by intracellular signaling pathways. All experimental data of cell migration in this subsection is obtained from the work by Roman *et al*. [33], who recently reported a novel protein, Arpin. The Arpin antagonizes an intrinsic positive feedback loop sustaining lamellipodial protrusion, which promotes turning during cell migration, i.e., Arpin decreases the directional persistence of cell migration [8]. Roman *et al*. analyzed the migratory dynamics of three different mammalian cell types regulated by Arpin, including the mammary carcinoma cell line MDA-MB-231, the motile amoeba Dictyostelium discoideum (*D. discoideum*) and fish-scale keratocytes (FK). In particular, the latter two cells have been analyzed with our approach, the data includes *D. discoideum* cell trajectories for wild-type (WT), Arpin knockout (KO) and GFP-Arpin (GFP), and trajectories of FK cells microinjected with either buffer (Buffer) or full-length Arpin (FL). The corresponding results are shown in Figs. 4 and 5.

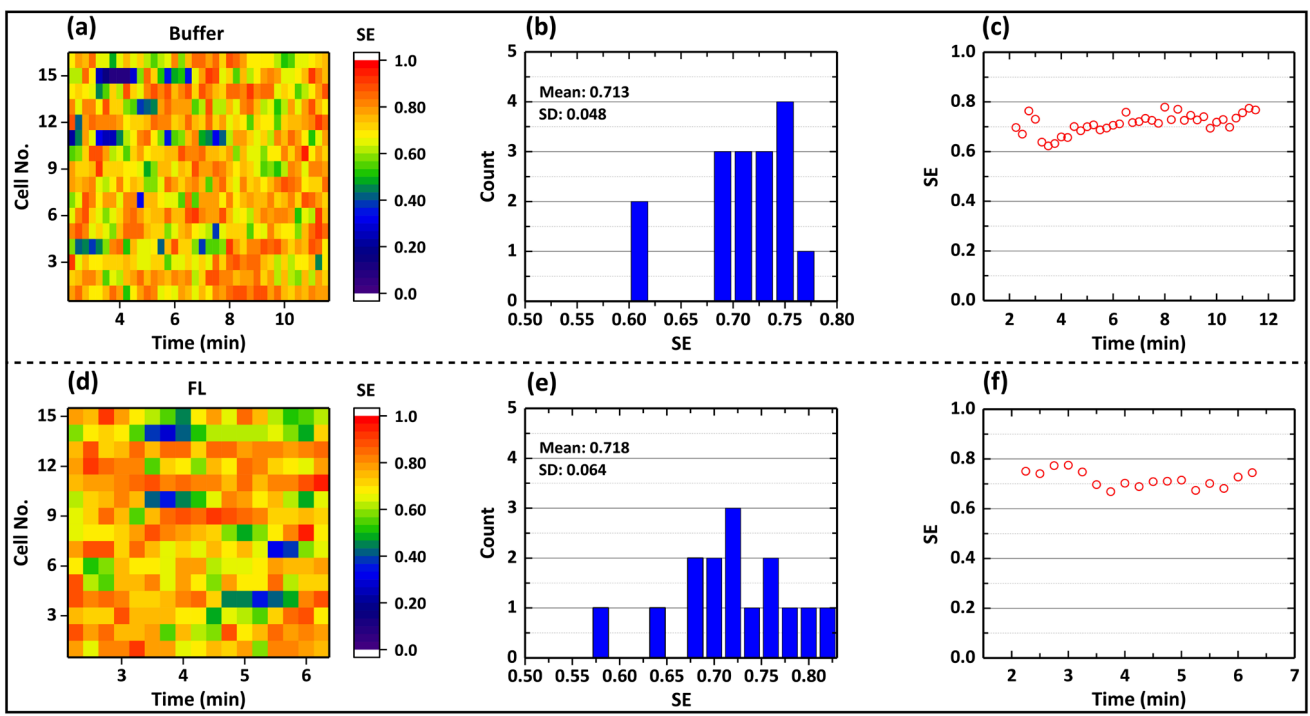

**Fig. 5.** Shannon entropy of the manually tracked fish keratocytes cells. The captions are same as those in Fig. 4, but the results corresponds to the cells microinjected with buffer (Buffer) (a-c) and full-length Arpin (FL) (d-f), respectively. Here, the time interval between frames is 15 s and the number of cells are 16 and 15 for Buffer and FL. All the experiment data is also obtained from Ref. [33].

Figures 4 and 5 show the time-varying SEs, the distributions of time-averaged SEs and cell-averaged SEs for *D. discoideum* cells with three different control groups and FK cells with two control groups. In particular, the cell-averaged SEs in Figs. 4(c, f, i) and Figs. 5(c, f) show greater fluctuations over time, which estimates unstable migration capabilities, and further reflects the unstable effects of ICSP to some extent.

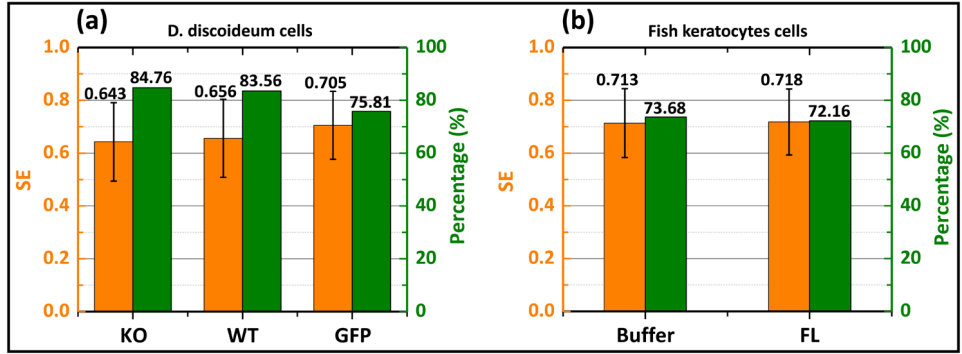

**Fig. 6.** The overall-averaged Shannon entropy and the percentage of entropies less than 0.8. The captions are same as those in Fig. 3(l), but the results corresponds to the *D. discoideum* cells (a) and fish keratocytes cells (b).

Similarly, we use average of all SEs and percentage of the SEs less than 0.8 to analyze the overall characteristics for cell populations, as shown in Fig. 6. The overall-averaged SE for KO is 0.643, which is less than 0.656 for WT and 0.705 for GFP, and the corresponding percentages are 84.76%, 83.56% and 75.81%, as seen in Fig. 6(a). Here, we believe the overall-averaged SE for KO should be smaller than 0.64, because the value 0.64 has been improved by the abnormal part [see red part around 10.5 min in Fig. 4(a), or the peak at 10.5 min in Fig. 4(c)]. Likewise, the FK cells microinjected with buffer exhibit smaller overall-averaged SE (0.713) and greater percentage (73.68%), when compared with the overall-averaged SE (0.718) and percentage (72.16%) for FL. These results further validate the protein, Arpin, does decrease the persistence of cell migration.

*4.2. Extracellular matrix: collagen gel vs. solid polystyrene substrate*

To further test the performance of the approach developed, we perform *in vitro* experiments to obtain migration dynamics of MCF-10A mammary epithelial cells on top of 3D collagen gel (about 2 mm thickness) and on 2D petri dish (solid polystyrene substrate) and compute the Shannon entropy (SE) to analyze these data sets. In particular, MCF-10A cells marked with green fluorescent protein (GFP) were obtained from China Infrastructure of Cell Line Resource. The culture medium of MCF-10A-GFP is Dulbecco's modified Eagle's medium-F12 (DMEM/F12, Corning) supplemented with 5% horse serum (Gibco), 1% penicillin/streptomycin (Corning), 20 ng/ml human EGF (Gibco), 10 μg/ml insulin (Roche Diagnostics Gmbh), 100 ng/ml cholera toxin (Sigma-Aldrich), and 0.5 μg/ml hydrocortisone (Sigma-Aldrich). Type I collagen extracted from rat tail tendon (Corning) was diluted and neutralized pH to approximately 7.2, then the collagen solution was spread on the substrate of petri dish and incubated in 37°C for 30 min till polymerized into 3D matrix with a thickness of around 2 mm. The final collagen concentration was 2 mg/ml for the tests. The cell suspension covered the matrix and stayed in the cell incubator overnight before imaging. For cell migration test, 0.5 μl of cell suspension with different concentrations of cells were dropped on top of collagen gel or solid petri dish, then incubated for 2 hours before imaging. Time-lapse images were obtained using both a confocal laser scanning microscope (CLSM) with a 25X water immersion objective, and an automatic inverted fluorescent microscope (Nikon Ti-E) with a 10X objective. Both microscopes are equipped with an on-stage cell-culture incubator to provide constant temperature of 37°C with humid 5% $CO_2$.

We obtain *in vitro* migration trajectories of MCF-10A cells on 3D collagen I hydrogel with a collagen concentration 2 mg/ml and thickness of approximately 2 mm by randomly distributing the MCF-10A cells on collagen-based ECM with a low cell density $10^4$ cells/cm$^2$. Further, we carry out control experiments by obtaining migration trajectories of MCF-10A cells on 2D petri dish by initially randomly distributing the cells on solid Polystyrene substrate with a number density $10^4$ cells/cm$^2$. We record the migration trajectories for the two cases above in sampling time 2 min for every frame and obtain totally 120 frames, respectively. The corresponding trajectories are shown in Fig .7(a) and Fig. 8(a).

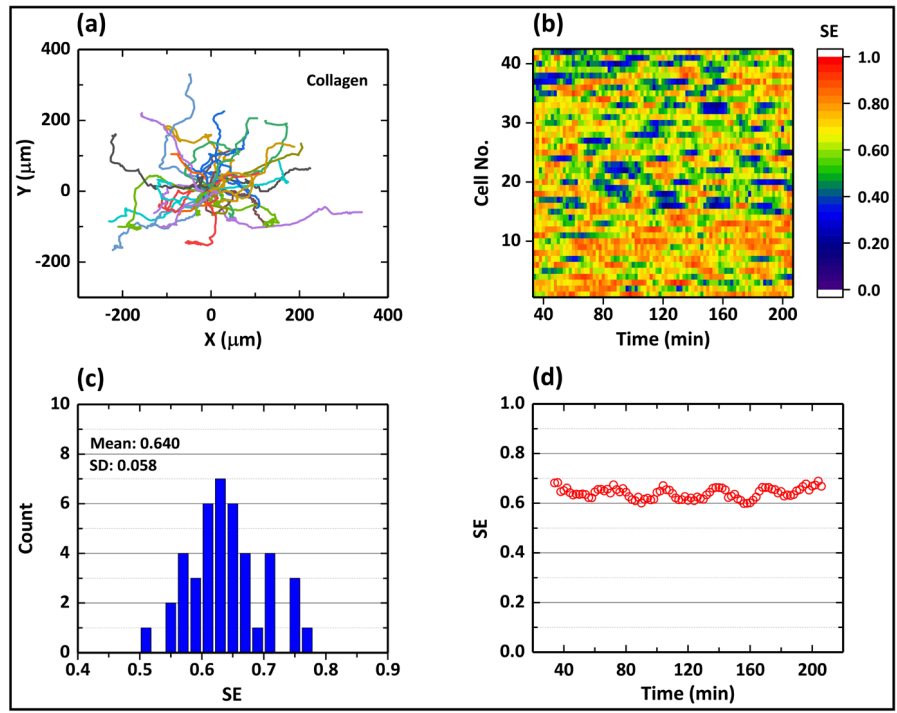

**Fig. 7.** Shannon entropy of the MCF-10A mammary epithelial cells on the top of 3D collagen gel. (a) All trajectories for 42 cells are moved to origin in x-y plane. Each color denotes individual trajectories. The other captions are same as those in Fig. 5.

The patterns in Fig. 7(a) and Fig. 8(a) are obtained by moving the start point of each trajectory to the origin in x-y plane, and which clearly show that the area explored by cells migrating on collagen layer is larger than that explored on 2D petri dish. We subsequently compute the time-varying Shannon entropies (SEs) for two cell groups above, as seen in Fig. 7(b) and Fig. 8(b). It is evident that the time-varying SEs for cells on collagen layer exhibit more small-SEs than those for cells on 2D petri dish. Moreover, both the time-varying SEs for the two groups indicate that some cells are associated with smaller SE, while other cells are associated with greater SE. Further, the time-averaged SE is computed for each cell group and their distributions are plotted in Fig. 7(c) and Fig. 8(c). The two distributions show that the overall-averaged SE (mean: 0.640, SD: 0.058) for cells on the collagen layer is less than that (mean: 0.689, SD: 0.096) for cells on 2D petri dish. When averaging the time-varying SEs along the vertical axis, we obtain the cell-averaged SEs, as seen in Fig. 7(d) and Fig. 8(d). The cell-averaged SE for cells on collagen layer almost locates in the interval [0.6, 0.7] and exhibits significant fluctuations, while the cell-averaged SE on 2D petri dish fluctuates around 0.7 with a smaller magnitude.

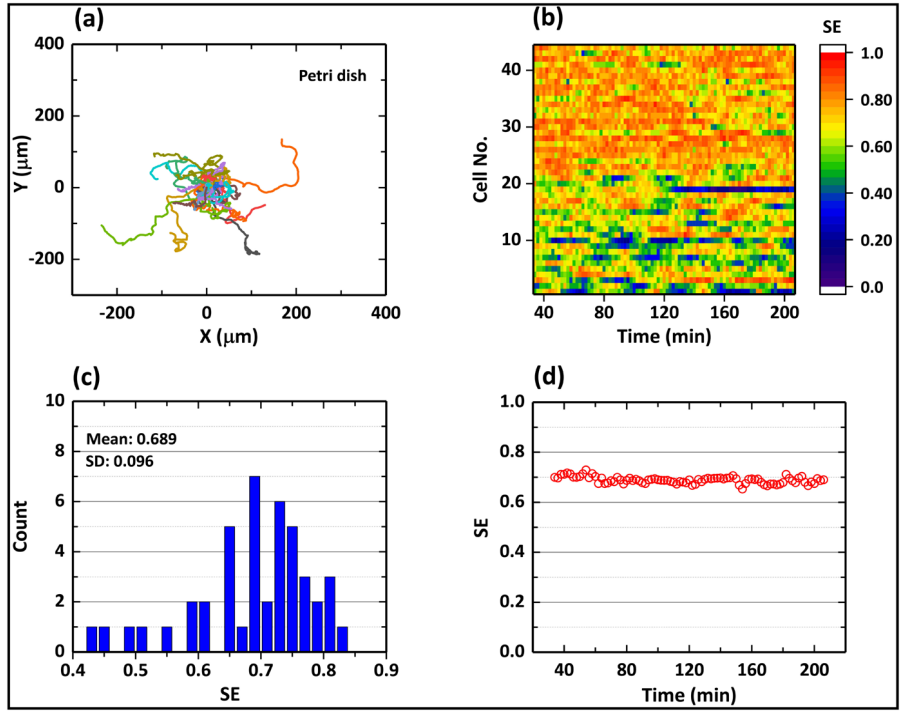

**Fig. 8.** Shannon entropy of the MCF-10A mammary epithelial cells on 2D petri dish. (a) All trajectories for 44 cells are moved to origin in x-y plane. Each color denotes individual trajectories. The other captions are same as those in Fig. 7.

In order to better compare the overall characteristics for the two cell groups, we also compute the average of all SEs and percentage of the SEs less than 0.8. Fig. 9 shows that the overall-averaged SE for cells on collagen layer (0.640) is less than that for cells on 2D petri dish (0.689), while the percentage of the former (84.95%) is greater than that of the latter (75.08%).

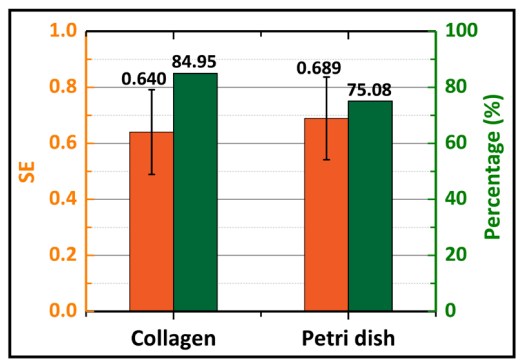

**Fig. 9.** The overall-averaged Shannon entropy and the percentage of entropies less than 0.8. The captions are same as those in Fig. 6, but the two left bars corresponds to the results for cells on the top of 3D collagen gel while the two right bars corresponds to those on 2D petri dish.

From the SE analysis, we can conclude that the migration of cells on collagen layer is more persistent than the cell migration on 2D petri dish. This is mainly because the fibrous microstructure of collagen gels can support long-range force propagation, which is crucial to mechanical signaling among the cells [10,11,56,57]. The strong motility of the MCF-10A cells can generate significant contractile forces during migration, and thus potentially induce strong cell-ECM mechanical coupling, which in turn inducing strong persistence of cell migration. However, the solid substrate in 2D petri dish is not able to transmit active cellular forces over long distances, thus no significant persistence in migration is expected.

5. Discussion and conclusions

In this paper, we introduce the time-varying Shannon entropy (SE) based on the wavelet power spectrum (WPS) obtained by performing wavelet transform (WT) of migration velocities and demonstrate its superior utility to characterize the persistence of cell migration. We first introduce a time-varying persistent random walk (TPRW) model and further construct the functions regarding motility parameters when considering the effects of intracellular signaling pathways (ICSP) or extracellular matrix (ECM), to simulate cell migration. On the basis of cell migration trajectories, we compute the velocity autocovariance function (VAC) and Fourier power spectrum (FPS), to quantitatively investigate the characteristics of cell migration in frequency domain. In order to study the effects of individual motility parameters on the shape of Fourier power spectrum, we define eight sets of parameters based on *control variable*. Results show the persistence time only increases the decay rate of Fourier power spectrum, while the migration speed does the amplitude. Moreover, inspired by the changes in Fourier power spectra, we then introduce Shannon entropy that can be obtained from the probabilities corresponding to power spectral values, to quantify the persistence of cell migration, i.e., the less the Shannon entropy, the closer the cell migration to a ballistic motion.

Due to the unstable cell migration capability, wavelet transform is employed to analyze cell migration velocities and derive wavelet power spectrum, which exhibits the time-frequency characteristics of cell migration. Because the local wavelet power spectrum along the frequency axis is identical with Fourier power spectrum of an OU process, we further compute the Shannon entropy for each local wavelet power spectrum, and naturally obtain the time-varying Shannon entropy, which reflects the time-varying persistence. In order to illustrate the utility and efficiency of our approach, we analyze trajectory data of *in vitro* cell migration regulated by distinct intracellular signaling pathways and extracellular matrix, exhibiting a rich spectrum of dynamic characteristics and persistence. In particular, our results based on the data of *D. discoideum* cell trajectories for wild-type (WT), Arpin knockout (KO) and GFP-Arpin (GFP), and trajectories of FK cells microinjected with either buffer (Buffer) or full-length Arpin (FL) indicate that the protein Arpin can decrease the persistence of cell migration. On the other hand, our analysis based on migration trajectories of MCF-10A cells on collagen layer and solid 2D substrate indicates that the collagen gel can significantly increase the persistence of cell migration. These observations are consistent with the findings reported in other literatures. We conclude by remarking that the Shannon entropy based on Fourier power spectrum can efficiently quantify the persistence of cell migration, while the Shannon entropy based on wavelet power spectrum can effectively capture the time-varying persistence of cell migration, which can also reflects the real-time effects of intracellular signaling pathways or extracellular matrix to some extent.

**Authors' contributions**

Yanping Liu: Conceptualization, Methodology, Writing - Original Draft. Yang Jiao: Methodology, Formal analysis. Qihui Fan: Investigation. Guoqiang Li: Resources. Jingru Yao: Visualization. Gao Wang: Software. Silong Lou: Validation. Guo Chen: Validation. Jianwei Shuai: Writing - Review & Editing. Liyu Liu: Supervision.

**Acknowledgments**

This research was supported by the National Natural Science Foundation of China (Grant Nos. 11974066, 11674043, 11675134, 11874310), the Fundamental Research Funds for the Central Universities (Grant No. 2019CDYGYB007), and the Natural Science Foundation of Chongqing, China (Grant No. cstc2019jcyj-msxmX0477, cstc2018jcyjA3679). Y. J. thanks Arizona State University for support during his sabbatical leave.

**References**

[1] M. Vicente-Manzanares and A. R. Horwitz, Methods in molecular biology (Clifton, N.J.) **769**, 1 (2011).

[2] P. Kulesa, D. L. Ellies, and P. A. Trainor, Developmental Dynamics **229**, 14 (2004).

[3] A. Tremel, A. Cai, N. Tirtaatmadja, B. D. Hughes, G. W. Stevens, K. A. Landman, and A. J. O'Connor, Chemical Engineering Science **64**, 247 (2009).

[4] G. D. Sharma, J. C. He, and H. E. P. Bazan, Journal of Biological Chemistry **278**, 21989 (2003).

[5] J. G. Cyster, Immunological Reviews **195**, 5 (2003).

[6] P. Friedl and S. Alexander, Cell **147**, 992 (2011).

[7] I. Dang *et al.*, Nature **503**, 281 (2013).

[8] R. Gorelik and A. Gautreau, Cytoskeleton **72**, 362 (2015).

[9] W. J. Polacheck, I. K. Zervantonakis, and R. D. Kamm, Cellular and Molecular Life Sciences **70**, 1335 (2013).

[10] H. Q. Nan, L. Liang, G. Chen, L. Y. Liu, R. C. Liu, and Y. Jiao, Physical Review E **97**, 13, 033311 (2018).

[11] Y. Zheng *et al.*, Physical Review E **100**, 13, 043303 (2019).

[12] J. Kim, Y. Zheng, A. A. Alobaidi, H. Q. Nan, J. X. Tian, Y. Jiao, and B. Sun, Biophysical Journal **118**, 1177 (2020).

[13] G. P. Gupta and J. Massague, Cell **127**, 679 (2006).

[14] A. Jemal, R. Siegel, J. Xu, and E. Ward, Ca-a Cancer Journal for Clinicians **60**, 277 (2010).

[15] E. A. Novikova, M. Raab, D. E. Discher, and C. Storm, Phys. Rev. Lett. **118**, 5, 078103 (2017).

[16] C. M. Lo, H. B. Wang, M. Dembo, and Y. L. Wang, Biophysical Journal **79**, 144 (2000).

[17] J. Park, D. H. Kim, H. N. Kim, C. J. Wang, M. K. Kwak, E. Hur, K. Y. Suh, S. S. An, and A. Levchenko, Nat. Mater. **15**, 792 (2016).

[18] J. Zhu, L. Liang, Y. Jiao, L. Liu, and U. S.-C. P. S.-O. Allianc, Plos One **10**, UNSP e0118058 (2015).

[19] W. Han *et al.*, Proceedings of the National Academy of Sciences of the United States of America **113**, 11208 (2016).

[20] E. A. Codling, M. J. Plank, and S. Benhamou, Journal of the Royal Society Interface **5**, 813 (2008).

[21] L. Li, S. F. Norrelykke, and E. C. Cox, Plos One **3**, 11, e2093 (2008).

[22] T. H. Harris *et al.*, Nature **486**, 545 (2012).

[23] G. H. Weiss, Physica a-Statistical Mechanics and Its Applications **311**, 381, Pii s0378-4371(02)00805-1 (2002).

[24] L. Li, E. C. Cox, and H. Flyvbjerg, Physical Biology **8**, 046006 (2011).

[25] Z. Sadjadi, M. R. Shaebani, H. Rieger, and L. Santen, Physical Review E **91**, 062715 (2015).

[26] D. S. Lemons and A. Gythiel, American Journal of Physics **65**, 1079 (1997).

[27] M. Schienbein and H. Gruler, Bulletin of Mathematical Biology **55**, 585 (1993).

[28] C. L. Stokes, D. A. Lauffenburger, and S. K. Williams, Journal of Cell Science **99**, 419 (1991).

[29] Y.-P. Liu, X.-C. Zhang, Y.-L. Wu, W. Liu, X. Li, R.-C. Liu, L.-Y. Liu, and J.-W. Shuai, Chinese Physics B **26**, 128707 (2017).

[30] Y.-P. Liu, X. Li, J. Qu, X.-J. Gao, Q.-Z. He, L.-Y. Liu, R.-C. Liu, and J.-W. Shuai, Frontiers of Physics **15**, 13602 (2020).

[31] C. L. Vestergaard, J. N. Pedersen, K. I. Mortensen, and H. Flyvbjerg, European Physical Journal-Special Topics **224**, 1151 (2015).

[32] C. L. Vestergaard, P. C. Blainey, and H. Flyvbjerg, Physical Review E **89**, 31, 022726 (2014).

[33] R. Gorelik and A. Gautreau, Nature Protocols **9**, 1931 (2014).

[34] J. N. Pedersen, L. Li, C. Gradinaru, R. H. Austin, E. C. Cox, and H. Flyvbjerg, Physical Review E **94**, 062401 (2016).

[35] P.-H. Wu, A. Giri, and D. Wirtz, Nature Protocols **10** (2015).

[36] A. J. Bergman and K. Zygourakis, Biomaterials **20**, 2235 (1999).

[37] C. Metzner, C. Mark, J. Steinwachs, L. Lautscham, F. Stadler, and B. Fabry, Nature Communications **6**, 8, 7516 (2015).

[38] P. Maiuri *et al.*, Cell **161**, 374 (2015).

[39] P.-H. Wu, A. Giri, S. X. Sun, and D. Wirtz, Proceedings of the National Academy of Sciences of the United States of America **111**, 3949 (2014).

[40] G. E. Uhlenbeck and L. S. Ornstein, Physical Review **36**, 0823 (1930).

[41] L. Cohen, Ieee Signal Processing Letters **5**, 292 (1998).

[42] N. Leibovich, A. Dechant, E. Lutz, and E. Barkai, Physical Review E **94**, 052130 (2016).

[43] C. E. Shannon, Bell Syst. Techn. J. **27**, 379 (1948).

[44] P. Kumar and E. Foufoula-Georgiou, Rev. Geophys. **35**, 385 (1997).

[45] G. Kaiser and L. H. J. P. T. Hudgins, **48**, 57 (1995).

[46] C. Torrence and G. P. Compo, Bull. Amer. Meteorol. Soc. **79**, 61 (1998).

[47] I. Daubechies, Computers in Physics **93**, 1671 (1992).

[48] K. M. Lau and H. Weng, Bull. Amer. Meteorol. Soc. **76**, 2391 (1995).

[49] J. Morlet, G. Arens, E. Fourgeau, and D. Giard, Geophysics **47**, 203 (1982).

[50] J. Morlet, G. Arens, E. Fourgeau, and D. Giard, Geophysics **47**, 222 (1982).

[51] P. Goupillaud, A. Grossmann, and J. Morlet, Geoexploration **23**, 85 (1984).

[52] A. Grossmann and J. Morlet, SIAM J. Math. Anal. **15**, 723 (1984).

[53] P. Flandrin, IEEE Trans. Inf. Theory **38**, 910 (1992).

[54] M. Farge, Annual Review of Fluid Mechanics **24**, 395 (1992).

[55] H. C. Shyu and Y. S. Sun, Multidimensional Systems and Signal Processing **13**, 101 (2002).

[56] L. Liang, C. Jones, S. H. Chen, B. Sun, and Y. Jiao, Physical Biology **13**, 11, 066001 (2016).

[57] H. Q. Nan, Y. Zheng, Y. H. H. Lin, S. H. Chen, C. Z. Eddy, J. X. Tian, W. X. Xu, B. Sun, and Y. Jiao, Soft Matter **15**, 6938 (2019).